\documentclass{ws-ijmpd}
\usepackage[super,compress]{cite}
\usepackage[dvips,breaklinks]{hyperref}
\hypersetup{colorlinks,urlcolor=black,citecolor=black,linkcolor=black,filecolor=black}

\begin{document}
\thispagestyle{plain}
\markboth{Jingbo Wang and Chao-Guang Huang}
{BF theory explanation of the entropy for rotating isolated horizons}

%
\catchline{}{}{}{}{}
%

\title{BF theory explanation of the entropy for rotating isolated horizons}

\author{Jingbo Wang}

\address{Institute of High Energy Physics and Theoretical Physics Center for
Science Facilities, \\ Chinese Academy of Sciences, Beijing, 100049, People's Republic of China,\\
wangjb@ihep.ac.cn}

\author{Chao-Guang Huang}

\address{Institute of High Energy Physics and Theoretical Physics Center for
Science Facilities, \\ Chinese Academy of Sciences, Beijing, 100049, People's Republic of China,\\
huangcg@ihep.ac.cn}
\maketitle

\begin{history}
\received{Day Month Year}
\revised{Day Month Year}
\end{history}

\begin{abstract}
In this paper, the isolated horizons with rotation are considered. It is shown that the symplectic form is the same as that in the nonrotating case. As a result, the boundary degrees of freedom can be also described by an SO$(1,1)$ BF theory. The entropy of the rotating isolated horizon satisfies the Bekenstein-Hawking area law with the same Barbero-Immirzi parameter.
\end{abstract}


\ccode{PACS numbers: 04.70.Dy,04.60.Pp}


\section{Introduction}
Isolated horizons (IHs) \cite{abf1,afk1} can be considered as generalization of the event horizon of the black hole. They have many applications in mathematical physics, numerical relativity and quantum gravity \cite{ak1}. Also they provide more physical setting for the entropy calculation in loop quantum gravity \cite{abck1,enpp1,smolin1}.
The IHs can be nonrotating or rotating. The calculation of the entropy was first did with the nonrotating case \cite{abk1}, and then generalized to the rotating case \cite{rot1,rot2}. Those calculation of the entropy are based on counting the dimension of
Hilbert spaces of the boundary U(1) Chern-Simons theory \cite{abk1}. Later, an SU(2) connection formulation is developed \cite{rot3}.

In previous paper Ref.~\refcite{wmz}, a new approach based on the BF theory to calculate the entropy of the isolated horizons is formulated. Unfortunately this method can only be applied to the nonrotating case then. In this paper, it will be shown that the symplectic form of the rotating isolated horizon is the same as that of the nonrotating case, thus the former case can be reduced to the later case, which has been dealt already.

The paper is organized as follows. In section 2, a brief review of the calculation of the entropy for nonrotating isolated horizon with BF theory is given. In section 3, it is shown that the symplectic form of the rotating isolated horizons can be reduced to the nonrotating case. So the same results of the entropy calculation can be achieved.  Section 4 is the conclusion.
\section{Review for the nonrotating case}
In this section, a brief review of the entropy calculation for the nonrotating isolated horizons based on SO$(1,1)$ BF theory is given \cite{wmz}. The starting point is the Palatini action of general relativity on 4-dimensional manifold $\mathcal{M}$,
\begin{equation}\label{1}
  S=-\frac{1}{2\kappa}\int_{\mathcal{M}} \Sigma_{IJ}   \wedge F^{IJ},
\end{equation}
where $\kappa\equiv 8\pi G$,
\begin{equation}\label{2}
\Sigma_{IJ}=\frac{1}{2}\varepsilon_{IJKL} e^K\wedge e^L,
\end{equation}
$e^I$ is the co-tetrad, $A^{IJ}$ is the SO$(3,1)$ connection 1-form, and $F(A)^{KL}\equiv {\rm d}A^{KL}+[A,A]^{KL}$ is the curvature of the connection $A^{KL}$. From this action, one can get the symplectic potential density and the symplectic current respectively. The isolated horizon is considered as the internal boundary of the spacetime, and it will contribute to the symplectic form.

Near the isolated horizon $\Delta$, the Bondi coordinates given by $(v,r,x^i,\,i=1,2)$ is adopted, where the horizon is given by $r=0$. The Newman-Penrose null co-tetrad $(l, n, m, \bar{m})$ can be written as \cite{ksh1}
\begin{equation} \label{2a}
\left\{ \begin{aligned}
         n &= -{\rm d}v, \\
         l &={\rm d}r-(\tilde{\kappa}r){\rm d}v-2 \mathbf{Re}(\pi^{(0)}\xi_i^{(0)})r {\rm d}x^i+\mathcal{O}(r^2), \\
m &=-\bar{\pi}^{(0)} r {\rm d}v+(1-\mu^{(0)} r)\xi_i^{(0)} {\rm d}x^i-(\bar{\lambda}^{(0)}r)\bar{\xi}_i^{(0)} {\rm d}x^i+\mathcal{O}(r^2),
        \end{aligned} \right.
\end{equation}
where $\mu,\pi,\lambda$ are the spin coefficients in the Newman-Penrose formalism \cite{np1}, $f^{(0)}$ means its value on $\Delta$ and $\tilde{\kappa}$ is the surface gravity on the horizon.

Next the following set of co-tetrad fields is chosen
\begin{equation}\begin{split}\label{2b}
    e^0=\sqrt{\frac{1}{2}}(\alpha n+\frac{1}{\alpha} l),\,e^1=\sqrt{\frac{1}{2}}(\alpha n-\frac{1}{\alpha} l),\\
    e^2=\sqrt{\frac{1}{2}}(m+\bar{m}),\, e^3=i\sqrt{\frac{1}{2}}(m-\bar{m}),
\end{split}\end{equation}
where $\alpha(x)$ is an arbitrary function of the coordinates. Each choice of $\alpha(x)$ characterizes a local Lorentz frame in the plane $\mathcal{I}$ formed by $\{e^0,e^1\}$.

Restricted to the horizon $\Delta$, the revelent co-tetrad fields (\ref{2b}) satisfy
\begin{equation}\label{3}\begin{split}
    e^0\triangleq e^1.
\end{split}\end{equation}
Hereafter the symbol $\triangleq$ denote equalities on $\Delta$.

After some straightforward calculation, the following conditions are obtained:
\begin{equation}\label{4}\begin{split}
\Sigma_{0i}\triangleq -\Sigma_{1i},\,A^{0i}\triangleq A^{1i},\forall i=2,3,\\
 A^{01}\triangleq \tilde{\kappa} {\rm d}v+{\rm d}(\ln\alpha)+\pi m+\bar{\pi} \bar{m}.
\end{split}\end{equation}
So the horizon integral of the symplectic current reduces to
\begin{equation}\label{5}
\Omega_{\Delta}(\delta_1,\delta_2)=\frac{1}{\kappa} \int_{\Delta}\delta_{[1}\Sigma_{IJ}\wedge \delta_{2]} A^{IJ}=\frac{2}{\kappa}\int_{\Delta} \delta_{[1}\Sigma_{01}\wedge\delta_{2]} A^{01}.
\end{equation}

For the nonrotating isolated horizons, the condition $\pi=0$ is satisfied, so
 \begin{equation}\label{6}
 {\rm d} A^{01}\triangleq 0.
\end{equation}
Also it is easy to see
\begin{equation}\label{7}
    {\rm d}\Sigma_{01}\triangleq0.
\end{equation}
Then an 1-form $B$ can be introduced locally on $\Delta$ to satisfy
\begin{equation}\label{8}
    {\rm d}B\triangleq\frac{\Sigma_{01}}{\kappa}.
\end{equation}The $B$ field has to satisfy the constraint
\begin{equation}\label{9}
    \oint_{H} {\rm d}B=\frac{1}{\kappa}\oint_{H} \Sigma_{01}=\frac{a_H}{\kappa},
\end{equation}
where $H$ is the spatial section of $\Delta$ and $a_H$ denotes the area of the horizon.

From above conditions, it can be shown that the integral (\ref{5}) can be written as \cite{wmz}
\begin{equation}\label{10}
\frac{2}{\kappa}\int_{\Delta} \delta_{[2}\Sigma_{01}\wedge\delta_{1]} A^{01}\triangleq
2\int_{H_2} \delta_{[2}B\wedge \delta_{1]}A^{01}-
2\int_{H_1} \delta_{[2}B\wedge \delta_{1]}A^{01},
\end{equation}
which is the symplectic form for a SO$(1,1)$ BF theory.

Above results suggest one to consider the system described by general relativity for the bulk ${\cal M}$ plus
an SO$(1,1)$ BF theory for the boundary $\Delta$.
In loop quantum gravity approach, only the horizon degrees of freedom contribute
to the IH entropy. Hence, the bulk degrees of freedom need to be traced out.

The bulk and boundary theories should be quantized separately. The full Hilbert space is the tensor product of those two sub-system $\mathcal{H}=\mathcal{H}_M\otimes \mathcal{H}_H$, where $\mathcal{H}_M$ is the Hilbert space for loop quantum gravity on the bulk, and $\mathcal{H}_H$ is the Hilbert space for the BF theory on the boundary. Then after applying the boundary condition, which is just the quantized version of the condition (\ref{8}), the relation of quantum number between bulk and boundary theories can be established. Finally the number of independent boundary quantum states which satisfy the constraint (\ref{9}) will give the entropy for the nonrotating isolated horizon \cite{wmz}:
\begin{equation}\label{11}
    S=\ln\mathcal{N}=\frac{\ln3}{\pi\gamma}\frac{a_H}{4l^2_{Pl}}+\ln \frac{2}{3},
\end{equation}
which gives the right answer if the Barbero-Immirzi parameter is set to be $\gamma=\ln 3/\pi$.
\section{Rotating isolated horizons}
For the rotating isolated horizon $\pi\neq 0$, the connection $A^{01}$ satisfy
\begin{equation}\label{12}
    {\rm d} A^{01}\triangleq 2\mathbf{Im}(\Psi_2) \Sigma_{01}\neq 0,
\end{equation}
where $\Psi_2$ is the second components of the Weyl tensor. This is the main obstruction to apply the above method. In this section, it will be shown that the rotating part doesn't affect the symplectic form, so our method can be applied to the rotating case either.

Eq.(\ref{4}) shows that the connection $A^{01}$ can be divided into two parts
\begin{equation}\label{13}
    A^{01}=\bar{A}^{01}+\tilde{A}^{01},
\end{equation}
such that
\begin{equation}\label{14}\begin{split}
   & \bar A^{01} \triangleq \tilde{\kappa} {\rm d}v+{\rm d}(\ln\alpha)\qquad {\rm with} \qquad
   {\rm d}\bar A^{01} \triangleq 0,\\
   & \tilde A^{01} \triangleq \pi m+\bar{\pi} \bar{m} .
   \end{split}
\end{equation}
Under the SO$(1,1)$ transformation which characterized by the $\alpha$, the $\tilde{A}^{01}$ is unchanged and $\bar{A}^{01}$ transforms as a connection.
Then, the horizon integral (\ref{5}) can be written as
\begin{equation}\label{15}
\Omega_{\Delta}(\delta_1,\delta_2)=\frac{2}{\kappa}\int_{\Delta} \delta_{[1}\Sigma_{01}\wedge\delta_{2]} \bar{A}^{01}+\frac{2}{\kappa}\int_{\Delta} \delta_{[1}\Sigma_{01}\wedge\delta_{2]} \tilde{A}^{01}.
\end{equation}
The first term looks like the symplectic form for a nonrotating IH.  Now, the key task is
to calculate
\begin{equation}\label{16}
  \Omega'_{\Delta}(\delta_1,\delta_2)=\frac{2}{\kappa}\int_{\Delta} \delta_{[1}\Sigma_{01}\wedge\delta_{2]} \tilde{A}^{01}.
\end{equation}

Following the method in Ref.~\refcite{enpp1}, we consider the variations $\delta=(\delta \Sigma_{01}, \delta \tilde{A}^{01})$ which, on the IH, correspond to linear combinations of SO$(1,1)$ internal gauge transformations and diffeomorphisms preserving the preferred foliation of $\Delta$.  That is, $\delta f$ on the IH can be written as
\begin{equation}\label{17}
    \delta f=\delta_{\gamma}f+\delta_w f,
\end{equation}
where $\gamma:\Delta\rightarrow \frak{so}(1,1)$ and $w:\Delta\rightarrow T(H)$. $\frak{so}(1,1)$ is the Lie algebra of SO$(1,1)$ and $T(H)$ is the tangent bundle over the spatial section $H$ of $\Delta$.  Since
both $\Sigma_{01}$ and $\tilde{A}^{01}$ are invariant under the SO$(1,1)$ transformation,
\begin{equation}\label{18}
    \Omega'_{\Delta}(\delta_{\gamma},\delta)=0.
\end{equation}
So we just need to consider two diffeomorphism transformations associated with two vector fields $w$ and $u$.
For diffeomorphism transformations associated with $w$, one has
\begin{equation}\label{19}\begin{split}
   & \delta_w \Sigma_{01}=\mathcal{L}_w \Sigma_{01}=w\lrcorner {\rm d} \Sigma_{01}+{\rm d} (w \lrcorner \Sigma_{01})={\rm d} (w \lrcorner \Sigma_{01}),\\
    & \delta_w \tilde{A}^{01}=\mathcal{L}_w \tilde{A}^{01}=w\lrcorner {\rm d} \tilde{A}^{01}+{\rm d} (w \lrcorner \tilde{A}^{01}),
\end{split}\end{equation}
where $w \lrcorner \Sigma_{01}$ is the interior product and ${\rm d}\Sigma_{01} \triangleq0$ due to the Einstein field equations.
So
\begin{eqnarray}\label{20}
   \Omega'_{\Delta}(\delta_w,\delta_u)
       &=&\frac{1}{\kappa}\int_{\Delta} {\rm d} (w \lrcorner \Sigma_{01})\wedge\delta_u\tilde{A}^{01}-\delta_u\Sigma_{01}\wedge(w\lrcorner {\rm d} \tilde{A}^{01})-\delta_u\Sigma_{01}\wedge {\rm d} (w \lrcorner \tilde{A}^{01}))\nonumber\\
    &=&\frac{1}{\kappa}\int_{\Delta} {\rm d} (w \lrcorner \Sigma_{01}\wedge\delta_u\tilde{A}^{01})+(w \lrcorner \Sigma_{01})\wedge\delta_u({\rm d}\tilde{A}^{01})-\delta_u\Sigma_{01}\wedge(w\lrcorner {\rm d} \tilde{A}^{01})\nonumber\\
    &&\qquad -{\rm d} (\delta_u\Sigma_{01}\wedge (w \lrcorner \tilde{A}^{01}))+\delta_u({\rm d} \Sigma_{01})\wedge(w\lrcorner \tilde{A}^{01})\nonumber\\
     &=&\frac{1}{\kappa}\int_{\partial \Delta} w \lrcorner \Sigma_{01}\wedge\delta_u\tilde{A}^{01}-\delta_u\Sigma_{01}\wedge (w \lrcorner \tilde{A}^{01}) \notag \\
    &&+\frac{1}{\kappa}\int_{\Delta}(w \lrcorner \Sigma_{01})\wedge\delta_u({\rm d} \tilde{A}^{01})-\delta_u\Sigma_{01}\wedge(w\lrcorner {\rm d} \tilde{A}^{01}).
 \end{eqnarray}
By using the identity $(w\lrcorner A_p)\wedge B_q+(-1)^p A_p\wedge (w\lrcorner B_q)=0$ for arbitrary $p$-form $A_p$,
 $q$-form $B_q$ and vector $w$ with $p+q-1$ equal to the dimension of the manifold,
Eq.(\ref{20}) reduces to
 \begin{eqnarray}\label{21}
   \Omega'_{\Delta}(\delta_w,\delta_u)&=&-\frac{1}{\kappa}\int_{\partial \Delta}\delta_u((w \lrcorner \tilde{A}^{01})\Sigma_{01})-\frac{1}{\kappa}\int_{\Delta}\delta_u(\Sigma_{01}\wedge(w\lrcorner {\rm d} \tilde{A}^{01})).
 \end{eqnarray}
The second term vanishes because of Eq.(\ref{12}).

For the first term, with the help of Eq.(\ref{19}) it becomes
\begin{eqnarray}\label{22}
   -\frac{1}{\kappa}\int_{\partial \Delta}\delta_u((w \lrcorner \tilde{A}^{01})\Sigma_{01})&=&-\frac{1}{\kappa}\int_{\partial \Delta}u\lrcorner {\rm d} ((w \lrcorner \tilde{A}^{01})\Sigma_{01})+{\rm d} (u\lrcorner((w \lrcorner \tilde{A}^{01})\Sigma_{01}))\nonumber\\=0,
 \end{eqnarray}
 since $\partial \Delta=H_2\cup (-H_1)$ is two dimension, and is also a boundary.

Thus, the symplectic form for the rotating isolated horizon Eq.(\ref{15}) is the same as that for nonrotating case,
\begin{equation}\label{25}
\Omega_{\Delta}(\delta_1,\delta_2)=\frac{2}{\kappa}\int_{\Delta} \delta_{[1}\Sigma_{01}\wedge\delta_{2]} \bar{A}^{01}.
\end{equation}

So the boundary degrees of freedom can also be described by an SO$(1,1)$ BF theory, and the method of calculating the entropy for a nonrotating IH in section 2 can be applied to rotating IHs. The derived entropy satisfies the Bekenstein-Hawking area law with the same Barbero-Immirzi parameter.

The above results can be generalized to the higher dimensional IHs straightly. The nonrotating IHs in $D$-dimensional spacetime have been dealt with in
Ref.~\refcite{wh1}.  The essential point for the
generalization to rotating IHs is to show that the expression in Eq.(\ref{16}) also vanishes in higher dimension.  It can be shown that the 01 component
of SO$(D-1,1)$ connection approaches to
\begin{equation}\label{26}
    A^{01}\triangleq (\tilde \kappa {\rm d} v+{\rm d}\ln \alpha)+(\delta^{\tt A B} \pi_{\tt A} e_{\tt B})  \triangleq : \bar A^{01}+ \tilde A^{01},
\end{equation}
on an IH, where $\tilde \kappa$ is still the surface gravity with respect to $l$, $\alpha$ is the
parameter of SO$(1,1)$ gauge group, $\{e_{\tt A}, A=2,\cdots, D-1\}$ is an orthogonal vielbein on a section of $\Delta$,
\begin{equation}\label{27}
    \pi_{\tt A} := e_{\tt A}^a l^b\nabla_b n_a,
\end{equation}
which is related to the angular momentum of the IH \cite{klp}.  For the internal gauge transformation, Eq.(\ref{18}) is still valid for higher dimension for the same reason.  For
diffeomorphic transformation generated by the vector field $w$ and $u$,
 \begin{eqnarray}\label{28}
   \Omega'_{\Delta}(\delta_w,\delta_u)&=&-\frac{2}{\kappa}\int_{\partial \Delta}\delta_u((w \lrcorner \tilde{A}^{01})\Sigma_{01})-\frac{2}{\kappa}\int_{\Delta}\delta_u(\Sigma_{01}\wedge(w\lrcorner {\rm d} \tilde{A}^{01})).
 \end{eqnarray}
The second term vanishes due to the fact that $\tilde{A}^{01}$ is independent of $v$
in the limit on $\Delta$.
For the same reason as in 4-dimensional case, the first term is zero since $\partial \Delta$ is a $D-2$ dimensional boundary.

\section{Conclusion}
The boundary BF theory approach to the statistical interpretation of the entropy can be applied to the rotating IHs.
The key reason is that Eq.(\ref{16}) does not contribute to
the symplectic form on the IH, though the component of
SO$(3,1)$ connection, $A^{01}$, acquires a nontrivial term $\tilde A^{01}$ related to the rotation.

The boundary BF theory can be applied to nonrotating and rotating isolated horizons in 4 dimensional and higher dimensional spacetime. It also can be applied to isolated horizons in Lovelock theory \cite{whl1}, so the range of the application is very wide.
\section*{Acknowledgments}
This work is supported by National Natural Science Foundation of China under the grant
11275207.
\bibliographystyle{ws-ijmpd}
\bibliography{btz8}
\end{document}